\RequirePackage{ifpdf}
\documentclass[11pt,a4paper]{article}
\usepackage{epsfig,jheppubme}

\usepackage{multirow,longtable,enumerate}

\usepackage{physics}
\usepackage{tensor}
\usepackage{amsmath}
\usepackage{amssymb}
\usepackage{appendix}
\usepackage{mathtools}
\usepackage{mathrsfs}
\usepackage[normalem]{ulem}
\usepackage{tikz}
\usepackage{booktabs}
\usepackage{bbm}
\usepackage{verbatim}
\usepackage{comment}
\usepackage{xspace}
\usepackage{longtable}

\usepackage{graphicx}
\usepackage{amsthm}
\usepackage{slashed}
\usepackage{hyperref}
\usepackage{setspace}
\usepackage{tikz}
\usepackage{array}
\usetikzlibrary{shapes}

\usepackage[tableposition=below]{caption}
\font\cmss=cmss12

\newcommand\half{\frac12}

\newcommand\bi{\begin{itemize}}
\newcommand\ei{\end{itemize}}

\newcommand\bea{\begin{eqnarray}}
\newcommand\eea{\end{eqnarray}}
\newcommand\be{\begin{equation}}
\newcommand\ee{\end{equation}}

\newcommand\bchi{{\overline \chi}}

\newcommand\ZZ{\hbox{Z\kern-.4emZ}}
\newcommand\sZZ{\hbox{\sevenfont Z\kern-.4emZ}}

\newcommand{\eref}[1]{Eq.\,(\ref{#1})}
\newcommand{\adst}{AdS$_3$\xspace}
\newcommand{\poin}{Poincar\'e\xspace}
\newcommand{\poinsum}{Poincar\'e sum\xspace}
\newcommand{\poinser}{Poincar\'e series\xspace}
\newcommand{\sutk}{SU(2)$_k$\xspace}

\newcommand{\sunk}{SU($N$)$_k$\xspace}
\newcommand{\sunkt}{SU($N$)$_k\,\times$ SU($N$)$_k$\xspace}
\newcommand{\sunone}{SU($N$)$_1$\xspace}
\newcommand{\sunonet}{SU($N$)$_1\,\times$ SU($N$)$_1$\xspace}
\newcommand\ninv{$N_{\rm inv}$}

\newcommand{\bc}{\mathbf{c}}

\allowdisplaybreaks

\def\IB{\relax{\rm I\kern-.18em B}}
\def\IC{{\relax\hbox{\kern.3em{\cmss I}$\kern-.4em{\rm C}$}}}
\def\ID{\relax{\rm I\kern-.18em D}}
\def\IE{\relax{\rm I\kern-.18em E}}
\def\IF{\relax{\rm I\kern-.18em F}}
\def\II{\relax{\rm I\kern-.18em I}}

\def\Id{\relax{1\kern-.32em 1}}
\def\IG{\relax\hbox{$\inbar\kern-.3em{\rm G}$}}
\def\IR{\relax{\rm I\kern-.18em R}}

\allowdisplaybreaks

\title{AdS$_3$ Gravity and RCFT Ensembles with Multiple Invariants} 

\author[a]{Viraj Meruliya\,}
\author[a]{and Sunil Mukhi\,}

\affiliation[a]{Indian Institute of Science Education and Research,\\  Homi Bhabha Rd, Pashan, Pune 411 008, India}

\emailAdd{viraj.meruliya29@gmail.com}
\emailAdd{sunil.mukhi@gmail.com}

\abstract{We use the Poincar\'e series method to compute gravity partition functions associated to SU($N$)$_1$ WZW models with arbitrarily large numbers of modular invariants. The result is an average over these invariants, with the weights being given by inverting a matrix whose size is of order the number of invariants. For the chosen models, this matrix takes a special form that allows us to invert it for arbitrary size and thereby explicitly calculate the weights of this average. For the identity seed we find that the weights are positive for {\em all} $N$, consistent with each model being dual to an ensemble average over CFT's.}
\preprint{}


\arxivnumber{}

\begin{document}

\maketitle

\section{Introduction and background}

The computation of AdS$_3$ gravity partition functions, under reasonable hypothesis about what quantum gravity should be, has led to some puzzles as well as illuminations. In the former category, the seminal work of Maloney-Witten \cite{Maloney:2007ud} provides a plausible method to compute the partition function for pure \adst gravity using the the method of \poinser. The computation requires regularisation at an intermediate stage, but one obtains a finite result at the end. The computation was streamlined and generalised in \cite{Keller:2014xba}. Unfortunately the result is inconsistent with our expectations for a quantum gravity partition function. The most serious problem is that in certain regions of parameter (energy/angular momentum) space, the degeneracies are generically negative \cite{Maloney:2007ud, Benjamin:2019stq, Keller:2014xba}. 

In view of this it becomes interesting to explore partition functions that correspond to generalisations of \adst gravity. One important step in this direction is the recent work in \cite{Maloney:2020nni, Afkhami-Jeddi:2020ezh}. Here the partition function has been calculated using the \poinser approach for topological U(1)$^{2D}$ Chern-Simons theories in $(2+1)$ dimensions. This is argued to be a gravity theory with gauge fields. The reasoning is that on the boundary there are U(1)$^{2D}$ Kac-Moody currents due to the CS action, as well as an energy-momentum tensor that is bilinear in these currents via the Sugawara construction. Thus the boundary degrees of freedom are precisely those expected from a bulk gauge and gravity theory. Meanwhile there no local bulk degrees of freedom because of the topological nature of the CS action, which is also what one expects for gravity in asymptotically \adst space-time. In these works it was found that the bulk calculation (performed using certain assumptions over what manifolds should be summed over) agrees with the average over a continuous family of boundary theories, performed using the Siegel-Weil measure on the space of Lorentzian (Narain) lattices. This is consistent with recent results in 2d gravity which indicate that the AdS/CFT can be realised in a form where an average must be taken over boundary CFT's \cite{Saad:2019lba, Witten:2020wvy}. Related ideas in the context of AdS$_3$ gravity were proposed in \cite{Cotler:2020ugk}.

A variation on the above theme was considered in \cite{Meruliya:2021utr}. Here the bulk CS action was taken to be non-Abelian, of the form \sunkt. The ``gravity'' partition function should then be defined by an appropriate sum over manifolds. To implement this, one performs a \poinsum over modular transforms of a ``seed'' partition function \footnote{A similar calculation was performed for Virasoro minimal models some time ago \cite{Castro:2011zq} and generalised to higher-genus boundaries \cite{Jian:2019ubz} and boundary CFT \cite{Karch:2020flx}.}.

On the general grounds summarised above, the boundary theory should then be a rational CFT with an \sunkt Kac-Moody algebra. It is known that given a set of Kac-Moody characters, there are in general multiple modular invariants that can be made out of them, that correspond to different RCFT -- all having the same central charge and Kac-Moody algebra \cite{Cappelli:1986hf, Cappelli:1987xt, Gannon:1992ty, Gannon:1994cf, Itzykson:1988hk, Degiovanni:1989ne, Bauer:1990xx, Schellekens:1992db}. One may then expect that a \poinser calculation of the gravity partition function would produce a linear combination of a discrete set of boundary CFT having the desired symmetries. 

This expectation has turned out to be correct in the cases considered in \cite{Meruliya:2021utr}, where particular attention was given to values of $N,k$ where there are two or at most three modular invariants. A complex picture emerged wherein for many (but not all) cases, the linear combinations of RCFT partition functions arise with non-negative coefficients, allowing us to interpret the result as a probabilistic average over a discrete ensemble of theories. However the limitation to two or three modular invariants is rather restrictive. Another limitation is that the results presented in \cite{Meruliya:2021utr} were for small values of the central charge.

In the present work we address these limitations by going beyond the examples considered in \cite{Meruliya:2021utr}. We work with the case where the bulk gravity and the boundary CFT have \sunonet Kac-Moody algebras for arbitrary $N$. For these theories the number of possible modular invariants varies with $N$ in a complicated way related to the prime divisors of $N$, but for large enough $N$ it can be arbitrarily large. Computing the weights of the ensemble average involves inverting a matrix of this order, so one might think it is a forbidding task in the case of multiple invariants. However we will find that the matrix to be inverted is the tensor product of very special symmetric Toeplitz matrices, each of which is readily inverted for any size. Thus we can explicitly compute the weights of each contributing CFT starting with the identity seed, and we find that they are non-negative for all $N$. The result is striking given that this was not the case for generic \sunk theories. We speculate that this special property is related to the fact that \sunone theories are all special points in the moduli space of U(1)$^{2D}$ lattice theories with $D=N-1$. 

\section{Modular invariants of SU($N$)$_{1}$ WZW models}

The \sunk WZW models form a large class of 2d rational CFT. The relevant details of \sunk WZW models are briefly reviewed in the Appendix. 
Generically they have several distinct modular invariants for every set of Kac-Moody characters. The classification of these modular invariants has been a subject of great interest. For $SU(2)_{k}$ it was carried out in \cite{Cappelli:1986hf, Cappelli:1987xt}. Subsequently it was completed for $SU(3)_{k}$ in \cite{Gannon:1992ty, Gannon:1994cf}. Finally, the classification for \sunone was provided in \cite{Itzykson:1988hk, Degiovanni:1989ne}. Some results for the general case of $SU(N)_{k}$ can be found in \cite{Bauer:1990xx}. 
	
For the rest of this paper we will specialise to $SU(N)_{1}$. Using (\ref{673}), we see that the allowed Dykin labels are given by $\lambda_{i}=\rho_{i}=1$ or $\lambda_{i}=\rho_{i} + \delta_{b,i}=1+\delta_{b,i}$ for some $b$ between $1$ to $N-1$. We denote these allowed Dynkin labels by $\lambda^{a}$ where $a\in\{0,1,\dots,N-1\}$ and $(\lambda^{a})_{i} = 1 + \delta_{a,i}$. The characters can now be labelled by the index $a$. It is then easy to check that the central charge and conformal dimensions are:
\be
\mathbf{c} = N-1,\qquad 
h_{a} = \frac{a(N-a)}{2N} 
\ee
and the $T$ and $S$ matrices for modular transformations (defined in generality in the Appendix) act as:  
\begin{align}
T_{a,a'} &= \delta_{a,a'}\exp(2\pi i \left( h_{a} - \frac{\bc}{24}\right) ) \\
S_{a,a'} &= \frac{1}{\sqrt{N}} \exp( -2\pi i \,\frac{aa'}{N} )
\end{align}

Using the results of \cite{Itzykson:1988hk, Degiovanni:1989ne}, we now list  the modular invariants that appear in these theories. First, define the positive integer $m$ by:
\begin{equation}
m = \begin{cases} N,~~  \mbox{ $N$ odd} \\  \frac{N}{2}, ~~ \mbox{ $N$ even} \end{cases}
\label{defofm}
\end{equation}
Then for every divisor $\delta$ of $m$, set $\alpha=[\delta,m/\delta]$ (where $[\,\,,\,]$ denotes the GCD) and $\omega(\delta) = \left(\rho\frac{m}{\alpha\delta}+\sigma\frac{\delta}{\alpha}\right) \mbox{ mod } \frac{N}{\alpha^{2}}$, where the integers $\rho,\sigma$ have been chosen to satisfy  $\rho\frac{m}{\alpha\delta}-\sigma\frac{\delta}{\alpha}=1$. Such $\rho,\sigma$ always exist because $\frac{m}{\alpha\delta}$ and $\frac{\delta}{\alpha}$ are coprime. They are unique up to shifts $(\rho,\sigma)\to (\rho,\sigma)+(\frac{\delta}{\alpha},\frac{m}{\alpha\delta})$ which do not affect $\omega(\delta)$. Now, note that $\omega(\frac{m}{\delta})=-\omega(\delta) \mbox{ mod }  \frac{N}{\alpha^{2}}$. 

Next we define the matrices:
\begin{equation}
\left( \Omega_{\delta} \right)_{a,a'} = \begin{cases} 0, \hspace{3.59cm} \alpha \nmid a \mbox{ or } \alpha \nmid a'    \\ \sum_{\xi=0}^{\alpha-1} \delta_{a', \, \omega(\delta)a + \xi \frac{N}{\alpha}}, \quad\quad \mbox{else} \end{cases}
\end{equation}
Then the following partition functions satisfy the properties of modular invariance, integrality and positivity, as well as non-degeneracy of the vacuum, and hence correspond to physical partition functions:
\begin{equation}
\label{683}
Z_{\delta} = \sum_{a=0}^{N-1}\bar{\chi}_{a} \left( \Omega_{\delta} \right)_{a,a'} \chi_{a'}
\end{equation}
Here the indices $a,a'$ in the matrix elements and characters will always be understood as integers modulo $N$. Note that the characters used here are the ``unflavoured'' ones, without a chemical potential for the currents in the Cartan sub-algebra.

For unflavoured characters $\{\chi_{a}\}$, the functions $\chi_{a}$ and $\chi_{N-a}$ $(a\ne0)$ are the same. This means that most of the characters have a twofold degeneracy, arising from the fact that a representation and its complex conjugate have the same character (there is one exception for all $N$, namely the identity character which under $a\to N-a$ is mapped to $\chi_N$ which then is the same as $\chi_0$ mod $N$. There is also another exception but only for even $N$ -- the character for $a=\frac{N}{2}$ is mapped to itself when $a\to N-a$). As a result, one finds that the partition functions satisfy:
\be
Z_{\delta}=Z_{\frac{m}{\delta}}
\label{Zsym}
\ee
This can be seen from:
\begin{equation}
\begin{split}
(\Omega_{\frac{m}{\delta}})_{a,a'} = \sum_{\xi=0}^{\alpha-1} \delta_{a', \, -\omega(\delta)a + \xi \frac{N}{\alpha}} = \sum_{\xi'=0}^{\alpha-1} \delta_{a', \, N - \omega(\delta)a - \xi' \frac{N}{\alpha}}  \quad (\alpha|a \mbox{ and } \alpha|a') \\
\end{split}
\end{equation}
where we used $\xi'=\alpha-\xi$. Then, 
\begin{equation}
\begin{split}
Z_{\frac{m}{\delta}} &= \sum_{\substack{a=0 \\ \alpha | a}}^{N-1} \sum_{\xi'=0}^{\alpha-1} \bar{\chi}_{a}\,  \chi_{N -\omega(\delta)a - \xi' N/\alpha} = \sum_{\substack{a=0 \\ \alpha | a}}^{N-1} \sum_{\xi'=0}^{\alpha-1} \bar{\chi}_{a}\,  \chi_{\omega(\delta)a + \xi' N/\alpha}= Z_{\delta} \\
\end{split}
\end{equation}
where we used $\chi_{a} = \chi_{N-a}$. Thus the number of linearly independent invariants is given by:
\begin{equation}
N_{\rm inv} = \begin{cases} \frac{\sigma(m)}{2}, \hspace{0.8cm} \sigma(m) \mbox{ even} \\ \frac{\sigma(m)+1}{2}, \quad \sigma(m) \mbox{ odd} \\ \end{cases}
\label{numberofZ}
\end{equation}
where $\sigma(m)=\sum_{d|m}1$ counts the divisors of $m$. However, we will often find it useful to consider the total list of invariants as $Z_\delta$ for all $\delta |m$, and impose the relation \eref{Zsym} only at the end.

Although the process of writing down the modular invariants of $SU(N)_{1}$ is very similar to that of $SU(2)_{k}$, they differ in the fact that for $SU(2)_{k}$ one can have at most three physical invariants, namely the A, D, E type while the remaining, if any, are unphysical. For $SU(N)_{1}$ there is no such restriction, and one can have an arbitrarily large number of physical modular invariants as we will see below. Moreover, all of them are physical in the sense that they have non-negative integer coefficients in their $q$-series after being correctly normalised. 

\section{The \poinsum for \sunone}

We perform the \poinsum as in \cite{Castro:2011zq, Meruliya:2021utr}, starting with the partition function for a seed primary and then summing over the coset $\Gamma'\backslash$PSL(2,Z) where $\Gamma'$ is a discrete finite-index subgroup of PSL(2,Z) that preserves the seed. The subgroup $\Gamma'$ need not be equal to $\Gamma_c$, which is the maximal subgroup that preserves the seed, since any finite-index subgroup of that will give the same answer up to an overall factor.

Due to the identifications between the partition functions as described in the previous Section, we adopt the following strategy for performing the Poincar\'e sum: at first we run the algorithm with the full set of characters, and only at the end impose the appropriate identifications between them. Likewise, when computing the coefficients of the resulting sum over physical modular invariants, we will use all the available $\Omega_{\delta}$ as the basis for a given model and then compute the cofficients $c_{\delta}$ corresponding to the modular invariant $Z_{\delta}$. At the end, we will impose the constraint that $Z_{\frac{m}{\delta}}=Z_{\delta}$. The case of two physical invariants has been dealt with in \cite{Meruliya:2021utr}. Here we consider first some classes of models whose number of invariants grows with $N$, and then move on to the most general case.

From the discussion in the previous section, the number of independent invariants \ninv\ is of order the divisor function $\sigma(m)$ where $m$ is either $N$ or $\frac{N}{2}$ (see \eref{defofm}). As was done for the simpler cases considered in \cite{Meruliya:2021utr}, we would now like to evaluate the \poinsum\ -- for cases with arbitrary  \ninv. This is equivalent to computing the coefficients of each invariant, which requires us to invert a $\sigma(m)\times\sigma(m)$ matrix of inner products. This appears to be a daunting task for arbitrarily large values of $\sigma(m)$. However, as we will see, the matrix in question has a special form that enables us to invert it for arbitrary $\sigma(m)$.

Since all of the modular invariants labelled by divisors of $m$ are physical, we do not need to restrict to any special sub-class of models as was required for $SU(2)_{k}$. We will see that for any given fixed number \ninv\ of invariants, there is in general an infinite family of models within this category. For each such family we can consider the large $\mathbf{c}$ limit. Therefore, in particular one can consider the large central charge limit for any desired number of physical invariants that one would like to study.

\subsection{Counting invariants}

To see what families lead to a given value of the total number of invariants, we start by writing the integer $m$ defined in \eref{defofm} in terms of its distinct prime factors:
\be
m=\prod_{i=1}^{s}p_{i}^{n_{i}}
\label{primedec}
\ee
Then the generalised divisor function $\sigma_{r}(m) \equiv \sum_{\delta|m}\delta^{r}$ is given by:
\begin{equation}
\sigma_{r}(m)  = \prod_{i=1}^{s}\sum_{j=0}^{n_{i}}p_{i}^{jr} = \prod_{i=1}^{s} \frac{p_{i}^{r(n_{i}+1)} - 1}{p_{i}^{n}-1}
\end{equation}
We are interested in the case $\sigma(m)\equiv\sigma_0(m)$, which counts the number of divisors. This can be found by taking the $r\rightarrow{0}$ limit in the above expression. This leads to:
\begin{equation}
\label{484}
\sigma(m) = \prod_{i=1}^{s}(n_{i}+1)
\end{equation}
This expression relates $\sigma(m)$ to the powers of the primes in $m$. For example, if we want to look at $\sigma(m)=6$, then we can write $6=2\times3$ or $6$. From \eref{primedec} this means that the allowed prime decompositions are $m=pq^{2}$ or $p^{5}$. This corresponds to SU($N$) for $N=pq^2,p^5, 2pq^2, 2p^5$ where $p,q$ are arbitrary distinct primes that moreover must be odd in the first two cases.

\subsection{A simple infinite sub-class}

From (\ref{484}), it is clear that $\sigma(m=p^{n}) = n+1$. So theories with $m=p^n$ form infinite families having a linearly growing number of physical invariants, namely $n+1$. We will start by discussing these and in the following subsection, move on to the most general case. 

For the above values of $m$ we actually have two classes of theories: $N=p^n$ for $p>2$ and $N=2p^n$ for $p\ge 2$. We will discuss each in turn. As before, we would like to know the coefficients of the multiple invariants in the general case. In particular it is important to check whether the Poincar\'e sum gives a result with only positive coefficients in the linear combination. We will focus on the vacuum (identity) seed in this paper, though our calculations contain the results for all seeds.

\subsubsection*{The sub-family $N=p^n$}

Let us start with the family $N=m=p^{n}$ where $p$ is a prime $>2$. The relevant matrices are $\Omega_{p^{k}}$ with $k\in\{0,1,2,\dots,n\}$. These are as follows:

\vspace{2mm}

\underline{Case I: $n$ even.} Here we have:
\begin{equation}
\begin{split}
\label{485}
&\hspace*{-2cm}k< \frac{n}{2}:\\[-2mm]
&(\Omega_{p^{k}})_{a,a'} = \begin{cases} 0, \hspace{3.4cm} p^{k}\nmid a \mbox{ or } p^{k}\nmid a'    \\ \sum_{\xi=0}^{p^{k}-1}\delta_{a',-a+\xi p^{n-k}}, \quad \mbox{else} \end{cases}\\[4mm]
&\hspace*{-2cm}k>\frac{n}{2}:\\[-2mm]
&(\Omega_{p^{k}})_{a,a'} = \begin{cases} 0, \hspace{3.2cm} p^{n-k}\nmid a \mbox{ or } p^{n-k}\nmid a'    \\ \sum_{\xi=0}^{p^{n-k}-1}\delta_{a',a+\xi p^{k}}, \quad \mbox{else} \end{cases}\\[4mm]
&\hspace*{-2cm}k=\frac{n}{2}:\\[-2mm]
&(\Omega_{p^{\frac{n}{2}}})_{a,a'} = \begin{cases} 0, \hspace{3cm} p^{\frac{n}{2}}\nmid a \mbox{ or } p^{\frac{n}{2}}\nmid a'    \\ \sum_{\xi=0}^{p^{\frac{n}{2}}-1}\delta_{a', \xi p^{\frac{n}{2}}}, \quad \mbox{else} \end{cases}
\end{split}
\end{equation}

In particular, the partition function corresponding to $\Omega_{p^{\frac{n}{2}}}$ is
\begin{equation}
Z_{p^{\frac{n}{2}}} = \sum_{\substack{a,a'=0\\p^{\frac{n}{2}}|a,a'}}^{N-1}\bar{\chi}_{a} \left( \sum_{\xi=0}^{p^{\frac{n}{2}}-1}\delta_{a', \xi p^{\frac{n}{2}}} \right) \chi_{a'} =  \sum_{\substack{a=0\\p^{\frac{n}{2}}|a}}^{N-1}\bar{\chi}_{a} \sum_{\xi=0}^{p^{\frac{n}{2}}-1} \chi_{ \xi p^{\frac{n}{2}}} 
\end{equation}
Now we note that:
\be
\sum_{\xi=0}^{p^{\frac{n}{2}}-1} \chi_{ \xi p^{\frac{n}{2}}} =\sum_{\substack{a=0\\p^{\frac{n}{2}}|a}}^{N-1}{\chi}_{a}
\ee
from which it follows that:
\be
Z_{p^{\frac{n}{2}}}= \abs{ \sum_{\substack{a=0\\p^{\frac{n}{2}}|a}}^{N-1}\bar{\chi}_{a}  }^{2}
\ee
This is the partition function of a holomorphically factorised CFT (the square of a meromorphic or one-character CFT). This is possible because the central charge $\bc=p^{n}-1$ is divisible by $24$ when $n$ is even and $p>3$. 

The matrix elements of the inner products can now be computed. For convenience, from now on we label the $\Omega_{p^{k}}$ as $\Omega_{k}$ where it will be understood that $k$ denotes the power of the prime $p$. The matrix elements of the inner product matrix are then labelled by $d_{k_{1}k_{2}} = \mbox{Tr}(\Omega_{k_{1}} \Omega_{k_{2}})$. This is a square matrix of rank $n+1$.

We start by looking at $d_{k_{1}k_{2}}$ where $k_{1}\le k_{2} < \frac{n}{2}$:
\begin{equation}
\label{489}
\begin{split}
d_{k_{1}k_{2}} &= \sum_{\substack{a,a'=0\\ p^{k_{2}}|a,a'}}^{N-1} \left( \sum_{\xi_{1}=0}^{p^{k_{1}}-1} \delta_{a', -a+\xi_{1}p^{n-k_{1}}} \right) \left( \sum_{\xi_{2}=0}^{p^{k_{2}}-1} \delta_{a, -a'+\xi_{2}p^{n-k_{2}}} \right) \\
&= \sum_{\substack{a=0\\ p^{k_{2}}|a}}^{N-1} \sum_{\xi_{1}=0}^{p^{k_{1}}-1}  \sum_{\xi_{2}=0}^{p^{k_{2}}-1}  \delta_{a, a - \xi_{1}p^{n-k_{1}} +\xi_{2}p^{n-k_{2}}} \\
\end{split}
\end{equation}
Now, the Kronecker delta selects the contribution for which 
$ - \xi_{1}p^{n-k_{1}} +\xi_{2}p^{n-k_{2}} = M p^{n}$ where $M\in\mathbb{Z}$. This is because the indices of the matrix are defined modulo $N=p^{n}$. This can be solved as $\xi_{2} = p^{k_{2}}(M + \xi_{1}p^{-k_{1}})$. Since, we have $0\le \xi_{2} < p^{k_{2}}$ and $0\le \xi_{1} < p^{k_{1}}$, we see that the only allowed value of $M$ is zero. Then for every allowed value of $\xi_{1}$, the value of $\xi_{2}$ is fixed to be $\xi_{1}p^{k_{2}-k_{1}}$. So (\ref{489}) reduces to:
\begin{equation}
\label{490}
\begin{split}
d_{k_{1}k_{2}} &= \sum_{\substack{a=0\\ p^{k_{2}}|a}}^{N-1} \sum_{\xi_{1}=0}^{p^{k_{1}}-1} 1 = p^{n-k_{2}+k_{1}} \\
\end{split}
\end{equation}
The result for $d_{k_{1}k_{2}} \, (k_{2}\le k_{1})$ follows from the cyclic property of trace -- it is simply $d_{k_{1}k_{2}} = p^{n-k_{1}+k_{2}}$. So, we have finally:
\begin{equation}
\label{491}
d_{k_1 k_2} = p^{n-\abs{k_{2}-k_{1}}} \quad (k_{1}, k_{2} < \frac{n}{2})
\end{equation}
A similar computation for the case where $k_1=\frac{n}{2}$ or $k_2=\frac{n}{2}$ or both, shows that the above expression is valid even in these cases. 

Let us now check the case of $d_{k_{1}k_{2}}$ with $\frac{n}{2}<k_{1}\le k_{2}$. Here we have:
\begin{equation}
\label{492}
\begin{split}
d_{k_{1}k_{2}} &= \sum_{\substack{a,a'=0\\ p^{n-k_{1}}|a,a'}}^{N-1} \left( \sum_{\xi_{1}=0}^{p^{n-k_{1}}-1} \delta_{a', a+\xi_{1}p^{k_{1}}} \right) \left( \sum_{\xi_{2}=0}^{p^{n-k_{2}}-1} \delta_{a, a'+\xi_{2}p^{k_{2}}} \right) \\
&= \sum_{\substack{a=0\\ p^{n-k_{1}}|a}}^{N-1} \sum_{\xi_{1}=0}^{p^{n-k_{1}}-1}  \sum_{\xi_{2}=0}^{p^{n-k_{2}}-1}  \delta_{a, a + \xi_{1}p^{k_{1}} +\xi_{2}p^{k_{2}}} \\
\end{split}
\end{equation}
The non-zero terms come from $ \xi_{1}p^{k_{1}} +\xi_{2}p^{k_{2}}= M p^{n} \,, M\in\mathbb{Z}$. We have then $\xi_{1} = p^{n-k_{1}}(M - \xi_{2}p^{-n+k_{2}})$. Using $0\le \xi_{2} < p^{n-k_{2}}$ and $0\le \xi_{1} < p^{n-k_{1}}$, we see that the allowed values of $M$ are $M=1$ for $0<\xi_{2}< p^{n-k_{2}}$ and $M=0$ for $\xi_{2}=0$. So (\ref{492}) reduces to:
\begin{equation}
\begin{split}
d_{k_{1}k_{2}} &= \sum_{\substack{a=0\\ p^{n-k_{1}}|a}}^{N-1} \sum_{\xi_{2}=0}^{p^{n-k_{2}}-1} 1  = p^{n-k_{2}+k_{1}} \\
\end{split}
\end{equation}
Thus the expression \eref{491} holds even in this case. As before, here we can set $k_{1}=\frac{n}{2}$ and the result is still valid. 

The only remaining case is $d_{k_{1}k_{2}}$ where $k_{1}<\frac{n}{2}<k_{2}$. Here,
\begin{equation}
\label{494}
\begin{split}
d_{k_{1}k_{2}} &= \sum_{\substack{a,a'=0\\ p^{l}|a,a'}}^{N-1} \left( \sum_{\xi_{1}=0}^{p^{k_{1}}-1} \delta_{a', -a+\xi_{1}p^{n-k_{1}}} \right) \left( \sum_{\xi_{2}=0}^{p^{n-k_{2}}-1} \delta_{a, a'+\xi_{2}p^{k_{2}}} \right) \\
&= \sum_{\substack{a=0\\ p^{l}|a}}^{N-1} \sum_{\xi_{1}=0}^{p^{k_{1}}-1}  \sum_{\xi_{2}=0}^{p^{n-k_{2}}-1}  \delta_{a, -a + \xi_{1}p^{n-k_{1}} + \xi_{2}p^{k_{2}}} \\
\end{split}
\end{equation}
where $l=$ max$(k_{1},n-k_{2})$. The non-zero terms come from values of $a$ which satisfy $2a = \xi_{1}p^{n-k_{1}} + \xi_{2}p^{k_{2}} + Mp^{n}\,, M\in\mathbb{Z}$. We can write $2a = p^{n}(\xi_{1}p^{-k_{1}} + \xi_{2}p^{-n+k_{2}} + M)$. Since, $0\le a<p^{n}$, $0\le \xi_{2} < p^{n-k_{2}}$ and $0\le \xi_{1} < p^{k_{1}}$ we see that the allowed values of $M$ are $0,\pm1$. Now, in the expression $2a = \xi_{1}p^{n-k_{1}} + \xi_{2}p^{k_{2}} + Mp^{n}$, if $\xi_{1}p^{n-k_{1}} + \xi_{2}p^{k_{2}}$ is even then $Mp^{n}$ must be even too. This is only possible when $M=0$. Similarly, if $(\xi_{1}p^{n-k_{1}} + \xi_{2}p^{k_{2}})$ is odd, then $M$ has to be either $1$ or $-1$. In this case, if $\xi_{1}p^{-k_{1}} + \xi_{2}p^{-n+k_{2}}<1$ then $M=1$ and if $\xi_{1}p^{-k_{1}} + \xi_{2}p^{-n+k_{2}}\ge1$ then $M=-1$. This is to ensure that  $0\le a<p^{n}$. Thus we see that for a given value of $\xi_{1}, \xi_{2}$, there is only one allowed value of $a$. So, if we perform the sum over $a$ in (\ref{494}), we just get $1$. This gives us:
\begin{equation}
d_{k_{1}k_{2}}  = \sum_{\xi_{1}=0}^{p^{k_{1}}-1}  \sum_{\xi_{2}=0}^{p^{n-k_{2}}-1} 1 = p^{n-k_{2}+k_{1}}
\end{equation}

With this result, we have proved that the matrix elements $d_{k_{1}k_{2}}$ are given by a simple relation:
\begin{equation}
\label{496}
d_{k_{1}k_{2}} = p^{n-\abs{k_{2}-k_{1}}} \quad (k_{1},k_{2} \in \{0,1,\dots,n\})
\end{equation}

\vspace{2mm}

\underline{Case II: $n$ odd.} In this case, we only have the matrices in the first two lines of (\ref{485}). There is no $\Omega_{\frac{n}{2}}$. The matrix elements $d_{k_{1}k_{2}}$ are then given by the same expression (\ref{496}).

To summarise, we have found that the matrix $d_{k_1k_2}$ takes the form of \eref{496} for all $N=p^n$. This is a very special matrix, being symmetric and Toeplitz (the latter property means the $ij$'th entry depends only on $i-j$). In fact it is even more special since every element is a power of $p^{-1}$ (up to an overall $p^n$ that we can take outside). We now want to invert this matrix to find the linear combination of modular invariants appearing in the Poincar\'e sum. 

It is easily verified that the inverse of the matrix $d_{k_{1}k_{2}}$ in \eref{496} above is given by:
\begin{equation}
\label{497}
d^{-1}_{k_{1}k_{2}} = \frac{1}{p^{n}(p^{2}-1)}
\begin{cases}
p^{2} + 1 - \delta_{k_{1}0} - \delta_{k_{1}n},\quad k_{1}=k_{2}  \\
-p,                                                     \hspace{3.3cm} \abs{k_{1}-k_{2}}=1  \\
0,                                                      \hspace{3.65cm} \mbox{else}
\end{cases}
\end{equation}

As in \cite{Meruliya:2021utr} (which the reader may consult for more details) we denote the seed partition function by $\bchi\, X_{\rm seed}\,\chi$. If we define $a_{k}=\mbox{Tr}(\Omega_{k}X_{\rm seed})$, then we know that for the vacuum seed $(X_{\rm seed})_{a,a'}=\delta_{a,0}\delta_{a',0}$ we have $a_{k}=1$ for all $k$. So, the Poincar\'e sum for the vacuum seed is proportional to
\begin{equation}
\label{4100}
\begin{split}
&(p^{2}-p)Z_{0} + (p-1)^{2}Z_{1} + (p-1)^{2}Z_{2} + \dots + (p-1)^{2}Z_{n-1} + (p^{2}-p)Z_{n} \\
&\quad = (p-1)\left( pZ_{0} + (p-1)Z_{1} + (p-1)Z_{2} + \dots + (p-1)Z_{n-1} + pZ_{n}  \right) \\
&\quad = 
\begin{cases} 
2(p-1)\left( pZ_{n} + (p-1)(Z_{n-1} + Z_{n-2} + \dots + Z_{\frac{n+1}{2}}) \right), \hspace{2.3cm} n \mbox{ odd} \\ 
2(p-1)\left( pZ_{n} + (p-1)(Z_{n-1} + Z_{n-2} + \dots + Z_{\frac{n}{2}+1}) + \frac{(p-1)}{2}Z_{\frac{n}{2}} \right), \hspace{0.3cm} n \mbox{ even}
\end{cases}\\
\end{split}
\end{equation}
where in the last line we have used the identifications $Z_{k}=Z_{n-k}$. This is an encouraging result: all the coefficients are positive and hence this expression can be interpreted as being proportional to the probabilities for averaging over different boundary RCFT's. This was shown to be true in \cite{Meruliya:2021utr} for many cases involving two or three invariants, but here we see that it holds in far greater generality, even with an arbitrarily large number of invariants. 

\subsubsection*{The sub-family $N=2p^n$}

Let us now more briefly discuss the case of even $N$, namely $N=2p^{n}$ where $(p\ge 2)$. The computations are very similar to the previous case of $N=p^{n}$, but with extra factors of $2$ at every step. Since $m=\frac{N}{2}=p^{n}$, the number of $\Omega_{k}$ is the same. They are given by:
\begin{equation}
\begin{split}
\label{4101}
&\hspace{-2cm}k< \frac{n}{2}: \\[-2mm]
&(\Omega_{k})_{a,a'} = \begin{cases} 0, \hspace{3.4cm} p^{k}\nmid a \mbox{ or } p^{k}\nmid a'    \\ \sum_{\xi=0}^{p^{k}-1}\delta_{a',-a+2\xi p^{n-k}}, \quad \mbox{else} \end{cases}\\[4mm]
& \hspace{-2cm}k>\frac{n}{2}:\\[-2mm]
&(\Omega_{k})_{a,a'} = \begin{cases} 0, \hspace{3.2cm} p^{n-k}\nmid a \mbox{ or } p^{n-k}\nmid a'    \\ \sum_{\xi=0}^{p^{n-k}-1}\delta_{a',a+2\xi p^{k}}, \quad \mbox{else} \end{cases}
\end{split}
\end{equation}
If $n$ is even we also have the matrix for $k=\frac{n}{2}$:
\begin{equation}
\label{4103}
(\Omega_{\frac{n}{2}})_{a,a'} = \begin{cases} 0, \hspace{3cm} p^{\frac{n}{2}}\nmid a \mbox{ or } p^{\frac{n}{2}}\nmid a'    \\ \sum_{\xi=0}^{p^{\frac{n}{2}}-1}\delta_{a', a+2\xi p^{\frac{n}{2}}}, \quad \mbox{else} \end{cases}
\end{equation}

Now we can compute the inner products between these matrices. The computation is essentially as before, however now the sum over $a$ has double the range and gives rise to an extra factor of two. For example, for $k_{1}\le k_{2}<\frac{n}{2}$ (analogous to the computaion in \eref{489}), we find: 
\begin{equation}
\label{4104}
\begin{split}
d_{k_{1}k_{2}} &= \sum_{\substack{a,a'=0\\ p^{k_{2}}|a,a'}}^{N-1} \left( \sum_{\xi_{1}=0}^{p^{k_{1}}-1} \delta_{a', -a+2\xi_{1}p^{n-k_{1}}} \right) \left( \sum_{\xi_{2}=0}^{p^{k_{2}}-1} \delta_{a, -a'+2\xi_{2}p^{n-k_{2}}} \right) \\
&= \sum_{\substack{a=0\\ p^{k_{2}}|a}}^{N-1} \sum_{\xi_{1}=0}^{p^{k_{1}}-1}  \sum_{\xi_{2}=0}^{p^{k_{2}}-1}  \delta_{a, a -2 \xi_{1}p^{n-k_{1}} + 2\xi_{2}p^{n-k_{2}}} \\
&= \sum_{\substack{a=0\\ p^{k_{2}}|a}}^{N-1} \sum_{\xi_{1}=0}^{p^{k_{1}}-1} 1 = 2p^{n-k_{2}+k_{1}} \\
\end{split}
\end{equation}
In going to the third line we have used the fact that the Kronecker delta gives non-zero terms when $ -2\xi_{1}p^{n-k_{1}} + 2\xi_{2}p^{n-k_{2}} = 2M p^{n} \,, M\in\mathbb{Z}$ which is exactly the same condition as before. The other computations follow similarly.

The computation for the case $k_{1}<\frac{n}{2}<k_{2}$ (analogous to the discussion below \eref{494}) is slightly different but again leads to the same conclusion. Here we have the condition that $a = \xi_{1}p^{n-k_{1}} + \xi_{2}p^{k_{2}} + Mp^{n}\,, M\in\mathbb{Z}$. Now, since $0\le a<2p^{n}$, $0\le \xi_{2} < p^{n-k_{2}}$ and $0\le \xi_{1} < p^{k_{1}}$, we have $M=0,\pm1$. If $\xi_{1}p^{-k_{1}} + \xi_{2}p^{-n+k_{2}}<1$ then $M=0$ or $1$ and if $\xi_{1}p^{-k_{1}} + \xi_{2}p^{-n+k_{2}}\ge 1$ then $M=0$ or $-1$. Thus for a given value of $\xi_{1},\xi_{2}$, we have two allowed values of $a$. The matrix element is then:
\begin{equation}
d_{k_{1}k_{2}} =   \sum_{\xi_{1}=0}^{p^{k_{1}}-1}  \sum_{\xi_{2}=0}^{p^{n-k_{2}}-1} 2 = 2p^{n-k_{2}+k_{1}}
\end{equation}
Thus we see that all the matrix elements are exactly the same, with just an overall factor of $2$. Then the inverse is still given by the expression (\ref{497}) with an extra $\half$ in front of it, and the Poincar\'e sum will give the same result as in \eref{4100} up to an irrelevant overall factor.

So far we have only considered specific sub-families of values of $N$, but we will now extend this to all values of $N\ge 2$. 

\subsection{The general case}

We now move on to the general case of arbitrary $m$. Recall that we defined its prime decomposition in \eref{primedec} above in terms of its distinct prime factors $p_i$.  Now, any divisor of $m$ is of the form $\delta=\prod_{i=1}^{s}p_{i}^{k_{i}}$ where $0\le k_{i}\le n_{i}$. From this it is easy to see that the total number of divisors are $\prod_{i=1}^{s}(n_{i}+1)$ confirming the result (\ref{484}). Thus from now on, we label the corresponding $\Omega_{\delta}$ as $\Omega_{\vec{k}}$ with $\vec{k}=(k_{1},\dots,k_{s})$. We want to compute the matrix elements $d_{\vec{k},\vec{l}} = \mbox{Tr}(\Omega_{\vec{k}}\Omega_{\vec{l}})$.

The procedure is similar to the simpler case of a single prime factor discussed above. Consider the entries $d_{\vec{k},\vec{l}}$ such that $k_{i}\le l_{i}\le \frac{n_{i}}{2}$. The $\Omega_{\delta}$ matrices in this case are:
\begin{equation}
\Omega_{\vec{k}} = 
\begin{cases} 0, \hspace{4.6cm} \prod_{i}p_{i}^{k_{i}} \nmid a \mbox{ or } \prod_{i}p_{i}^{k_{i}} \nmid a'    \\ \sum_{\xi=0}^{\prod_{i}p_{i}^{k_{i}}-1}\delta_{a',-a+\xi \prod_{i}p_{i}^{n_{i} - k_{i}}}, \quad \mbox{else} \end{cases}
\end{equation}
where we have  $\alpha=[\prod_{i}p_{i}^{k_{i}},\prod_{i}p_{i}^{n_{i}-k_{i}}]=\prod_{i}p_{i}^{k_{i}}$ and so $\omega=-1$ mod $N/\alpha^{2}$. Using this we compute $d_{\vec{k},\vec{l}}\,, (k_{i}\le l_{i}\le n_{i}/2)$ to find:
\begin{equation}
\begin{split}
d_{\vec{k},\vec{l}} &= \sum_{\substack{a,a'=0\\ \prod_{i}p_{i}^{l_{i}} | a,a'}}^{N-1} \sum_{\xi_{1}=0}^{\prod_{i}p_{i}^{k_{i}}-1}\delta_{a',-a+\xi_{1} \prod_{i}p_{i}^{n_{i} - k_{i}}} \sum_{\xi_{2}=0}^{\prod_{i}p_{i}^{l_{i}}-1}\delta_{a,-a'+\xi_{2} \prod_{i}p_{i}^{n_{i} - l_{i}}} \\
&=  \sum_{\substack{a=0\\ \prod_{i}p_{i}^{l_{i}} | a}}^{N-1} \sum_{\xi_{1}=0}^{\prod_{i}p_{i}^{k_{i}}-1} \sum_{\xi_{2}=0}^{\prod_{i}p_{i}^{l_{i}}-1}\delta_{a, a-\xi_{1} \prod_{i}p_{i}^{n_{i} - k_{i}}+\xi_{2} \prod_{i}p_{i}^{n_{i} - l_{i}}}  \\
\end{split}
\end{equation}
As before, the Kronecker delta gives a non-zero contribution when $ - \xi_{1}\prod_{i}p^{n_{i}-k_{i}} +\xi_{2}\prod_{i}p^{n_{i}-l_{i}} = M\prod_{i}p^{n_{i}}$ where $M\in\mathbb{Z}$. Due to the ranges of $\xi_{1},\xi_{2}$, we see that the only allowed value of $M$ is zero. Then for every allowed value of $\xi_{1}$, the value of $\xi_{2}$ is fixed to be $\xi_{1}\prod_{i}p^{l_{i}-k_{i}}$. We thus find:
\begin{equation}
\begin{split}
d_{\vec{k},\vec{l}} &= \sum_{\substack{a=0\\ \prod_{i}p_{i}^{l_{i}} | a}}^{N-1} \sum_{\xi_{1}=0}^{\prod_{i}p_{i}^{k_{i}}-1} 1 = \prod_{i} p_{i}^{n_{i}-l_{i}+k_{i}}
\end{split}
\end{equation}
The result is very similar to (\ref{496}) but now we have a contribution from each of the prime factors $p_{i}$. From these observations we conclude that for the general case:
\begin{equation}
\begin{split}
\label{4107}
d_{\vec{k},\vec{l}}~&= \prod_{i=1}^{s}p_{i}^{n_{i}-\abs{l_{i}-k_{i}}} ~~~~~\hbox{for } N (=m=\prod_{i}^{s}p_{i}^{n_{i}}) \hbox{ odd,}\\
&= 2\prod_{i=1}^{s}p_{i}^{n_{i}-\abs{l_{i}-k_{i}}} ~~~\hbox{for } N (=2m=2\prod_{i}^{s}p_{i}^{n_{i}}) \hbox{ even.}
\end{split}
\end{equation}
The previous result (\ref{496}) is a special case of (\ref{4107}) with $s=1$. The general result is a tensor product of the result for each prime factor. Writing $d_{\vec{k},\vec{l}} = \prod_{i=1}^{s}d_{k_{i},l_{i}}$, it is easy to find the inverse -- it is given by the expression $d^{-1}_{\vec{k},\vec{l}} = \prod_{i=1}^{s}d^{-1}_{k_{i},l_{i}}$ where $d^{-1}_{k_{i},l_{i}}$ has been computed in \eref{497}. 

To evaluate the Poincar\'e sum for the vacuum seed we use the fact that Tr$(\Omega_{\vec{k}}X_{\rm seed})=1$, so the relative coefficients appearing in front of the partition function $Z_{\vec{k}}$ are:
\begin{equation}
\label{4110}
C_{\vec{k}}= \sum_{\vec{l}}\prod_{i=1}^{s} d^{-1}_{k_{i},l_{i}}
\end{equation}
The sum over $\vec{l}$ is a sum over $l_{i}$ which lie in the range $0\le l_{i}\le n_{i}$. \eref{4110} can then be evaluated once we know the values $\sum_{l_{i}=0}^{n_{i}}d^{-1}_{k_{i},l_{i}}$. From the previous section we know that the entries of this quantity are equal to $p_i(p_i-1)$ if $k_{i}=0$ or $n_{i}$, else they are equal to $(p_i-1)^{2}$. From this we see the important fact that even in the general case, the weights appearing in the \poinsum are all positive. 

In fact we can write a generating function for the Poincar\'e sum of the vacuum seed:
\begin{equation}
\prod_{i=1}^{s}( p_i x_{i}^{0} + (p_i-1)x_{i}^{1} + \dots +  (p_i-1)x_{i}^{n_{i}-1} + p_i x_{i}^{n_{i}}) 
\end{equation}
The coefficient $C_{\vec{k}}$ is simply the coefficient of $x_{1}^{k_{1}}\dots x_{s}^{k_{s}}$ in this product. Once we have worked out the answer then we can, as usual, make the identification $Z_{\vec{k}}=Z_{\vec{n}-\vec{k}}$.

As an example, consider $m=\prod_{i=1}^{s}p_{i}$, i.e. each prime factor appears with multiplicity 1. Then we have $n_{i}=1$ $\forall i$, so the generating function is $\prod_{i=1}^{s}( p_i x_{i}^{0} + p_i x_{i}^{1})$. This in turn  means that $C_{\vec k}=1$ for all $\vec k$, up to an overall factor of $\prod_{i=1}^s p_i=m$. Thus, all possible partition functions are averaged with equal weights -- the Poincar\'e sum in this case is proportional to $\sum_{\vec{k}}Z_{\vec{k}}$.

In Table \ref{T8} we list the results of an explicit calculation of \poinsum (up to overall factors) for some selected examples with up to six invariants. These all confirm the general predictions above. 

\begin{longtable}{|>{\centering}p{0.8cm}|>{\centering}p{2cm}|>{\centering}p{0.8cm}|>{\centering}p{9.8cm}|}
\hline

 \hline
 $N$ & $m$ & $\sigma(m)$ & \textbf{Poincar\'{e} sum}   \tabularnewline
 \hline
 \hline

$24$ & $2^{2}\times3$ & 6 & $2Z_{(2,1)} + Z_{(1,1)} + 2Z_{(2,0)} $  \tabularnewline
\hline

$36$ & $2\times3^{2}$ & 6 & $3 Z_{(1,2)} + 3 Z_{(0,2)} + 2 Z_{(1,1)}$ \tabularnewline
\hline

$60$ & $2\times3\times5$ & 8 &  $Z_{(1,1,1)} + Z_{(0,1,1)} + Z_{(1,0,1)} + Z_{(1,1,0)}$\tabularnewline
\hline

$72$ & $2^{2}\times3^{2}$ & 9 &  $6Z_{(2,2)}+
3Z_{(1,2)}+4Z_{(2,1)}+6Z_{(0,2)}+Z_{(1,1)}$
\tabularnewline
\hline

$80$ & $2^{3}\times5$ & 8 & $2Z_{(3,1)}+
Z_{(2,1)}+Z_{(1,1)}+2Z_{(3,0)}$ \tabularnewline
\hline

$96$ & $2^{4}\times3$ & 10 &  $2Z_{(4,1)}+
Z_{(3,1)}+2Z_{(4,0)}+Z_{(2,1)}+Z_{(3,0)}$\tabularnewline
\hline

$105$ & $3\times5\times7$ & 8 & $Z_{(1,1,1)} + Z_{(0,1,1)} + Z_{(1,0,1)} + Z_{(1,1,0)}$  \tabularnewline
\hline

$120$ & $2^{2}\times3\times5$ & 12 &  $2Z_{(2,1,1)} + Z_{(1,1,1)} + 2Z_{(2,0,1)} + 2Z_{(0,1,1)} + 2Z_{(2,1,0)} + Z_{(1,0,1)}$
\tabularnewline
\hline

$144$ & $2^{3}\times3^{2}$ & 12 &  $6Z_{(3,2)} + 3Z_{(2,2)} + 4Z_{(3,1)} + 3Z_{(1,2)} + 2Z_{(2,1)} + 6Z_{(0,2)}$ \tabularnewline
\hline

$168$ & $2^{2}\times3\times7$ & 12 &  $2Z_{(2,1,1)} + Z_{(1,1,1)} + 2Z_{(2,0,1)} + 2Z_{(0,1,1)} + 2Z_{(2,1,0)} + Z_{(1,0,1)}$\tabularnewline
\hline
\caption{Poincar\'{e} Sums for some $SU(N)_{1}$ WZW models with multiple physical invariants.}
\label{T8}
\end{longtable}

\section{Discussion}

 The \poinsum for certain classes of rational CFT has been computed in  \cite{Castro:2011zq, Meruliya:2021utr, Benjamin:2021wzr}. In most of these cases, generically there appeared unphysical modular invariants (with negative ``degeneracies'') as well as negative coefficients multiplying physical modular invariants. For example \cite{Meruliya:2021utr} found infinite families of unitary RCFT corresponding to a sub-class of \sutk models with carefully chosen values of the level $k$, for which the \poinsum gave rise to physical modular invariants with non-negative coefficients, however for other values of the level one had unphysical results. In the present work we have found a more favourable situation -- for all \sunone current algebras, the \poinsum leads to an ensemble average of physical invariants with non-negative coefficients that we have calculated explicitly. Moreover these models are manifestly unitary. 
 
This adds an entire family's worth of data points to the set of examples where the \poinsum for 3d gravity ``works'', in the sense of giving a physically sensible ensemble average. The complementary set of examples where the sum ``fails to work'' includes many minimal models and \sutk models studied in \cite{Castro:2011zq, Meruliya:2021utr} as well as the much larger class of models considered in \cite{Benjamin:2021wzr} that include Calabi-Yau SCFT's. It also includes, importantly, the generic (irrational) case that was originally studied in \cite{Maloney:2007ud, Keller:2014xba}.
 
In \cite{Benjamin:2021wzr} it was conjectured that the \poinsum fails to work whenever $\bc>\bc_{\rm crit}$ -- the central charge of the gravity theory is greater than the critical value corresponding to the given boundary chiral algebra. Our result is consistent with this criterion. But the criterion does not tell us when the result should work, only when it should fail. It should be possible to find a stronger criterion that tells us which cases do work. The results presented here may be useful in finding this stronger criterion.
 
It is noteworthy that the \sunone WZW models, including all modular invariants, arise as specific Narain lattice \cite{Narain:1985jj, Narain:1986am} compactifications of $N-1$ free bosons, defined in terms of Lorentzian lattices $\Gamma_{N-1,N-1}$. The diagonal invariant arises from the Englert-Neveu lattice \cite{Englert:1985ws} while all others are discrete points in the space of Narain lattices. General lattices of this kind do not have enhanced non-Abelian symmetry, but form a continuous family with Abelian $U(1)^{2(N-1)}$ symmetry. As mentioned in the Introduction, the ensemble average over these has been studied in \cite{Afkhami-Jeddi:2020ezh, Maloney:2020nni}\footnote{Aspects of this problem that may be relevant to the present discussion have subsequently been investigated in \cite{Alday:2020qkm, Raeymaekers:2020gtz, Dymarsky:2020pzc, Datta:2021ftn}.}. The \poinsum obtained there coincides with the ensemble average using the Siegel-Weil measure on the space of lattices, which is a positive measure. The theories we discuss are discrete points in this space, and this seems to be the reason that we find the weights in the \poinsum to be positive. While these weights were derived here using the \poinsum on \sunone characters, it would be interesting to obtain them using a suitable restriction of the Siegel-Weil formula. Another interesting direction would be to relate our \poinsum to a calculation directly in the 3d \sunonet Chern-Simons theory that defines the bulk gravity, perhaps using the results of \cite{Porrati:2019knx}.

\section*{Acknowledgements}

We thank Alex Maloney and Palash Singh for helpful discussions. We are grateful to Ajay Salve and Kapil Ghadiali at TIFR Mumbai for their kind assistance in helping us generate some data. VM would like to acknowledge the INSPIRE Scholarship for Higher Education, Government of India. Finally, we are grateful for support from a grant by Precision Wires India Ltd.\ for String Theory and Quantum Gravity research at IISER Pune.

\appendices

\section{Some relevant details of \sunk theories}

From $N$ and the level $k$, we define the height $n=k+N$. Then the central charge of these families is:
    \begin{equation}
        \mathbf{c} = \frac{k(N^2-1)}{k+N} = (N^{2}-1) - \frac{N(N^{2}-1)}{n}
    \end{equation}
    For any given level $k$, only a finite number of representations are allowed. These are selected by the constraint:
    \begin{equation}
    \label{673}
        \sum_{i=1}^{N-1} \lambda_i < n
    \end{equation}
    where $\lambda\equiv(\lambda_1,\lambda_2,\dots,\lambda_{N-1})$ is the Dynkin label for $SU(N)$ \footnote{We are following the convention where the Dynkin label of the trivial representation is the unit vector $(1,1,\dots,1)$, instead of the zero vector.}, which labels the characters. The number of representations allowed by the constraint is given by:
    \begin{equation}
        \#_k = \frac{(N+k-1)!}{k!(N-1)!}
    \end{equation}
    The conformal dimension of the representation with Dynkin label $\lambda$ is:
    \begin{equation}
        h_\lambda = \frac{(\lambda-\rho,\lambda+\rho)}{2n}
    \end{equation}
    where $\rho=(1,1,\dots,1)$, and the inner product is $(x,y) = x_i\, \kappa_{i,j}\, y_j$, where $\kappa$ is the quadratic form matrix of $SU(N)$:
    \begin{equation}
        \kappa \equiv \frac{1}{N}
        \begin{pmatrix}
            N-1 & N-2 & N-3 & \cdots & 2 & 1 \\
            N-2 & 2(N-2) & 2(N-3) & \cdots & 4 & 2 \\
            N-3 & 2(N-3) & 3(N-3) & \cdots & 6 & 3 \\
            \vdots & \vdots & \vdots & \ddots & \vdots & \vdots \\
            2 & 4 & 6 & \cdots & 2(N-2) & N-2 \\
            1 & 2 & 3 & \cdots & N-2 & N-1 \\
        \end{pmatrix}
    \end{equation}
    The actions of the $S$ and $T$ modular transformations on the characters are:
	\begin{align}
		T_{\lambda\lambda'} &= \delta_{\lambda,\lambda'}\exp(2\pi i \left( h_{\lambda} - \frac{\mathbf{c}}{24}\right)) \\
		S_{\lambda\lambda'} &= \frac{i^{N(N-1)/2}}{\sqrt{Nn^{N-1}}} \sum_{w\in W} \mbox{det}(w)\, \exp(-2\pi i \left( \frac{\lambda \cdot w(\lambda')}{n} \right) )
	\end{align}
	where $W$ is the Weyl group of $SU(N)$.

\bibliographystyle{JHEP}
\bibliography{Multiple}

\end{document}